\documentstyle[preprint, aps]{revtex}

\begin{document}
\draft
\title{Classical calculation of high-order harmonic generation of atomic and
molecular gases in intense laser fields}
\author{{\small Chaohong Lee}$^{1\thanks{%
E-mail address: Chlee@wipm.whcnc.ac.cn.}}${\small ,Yiwu Duan}$^{2}${\small %
,Wing-Ki Liu}$^{3}${\small , Jian-Min Yuan}$^{4}${\small ,\ Lei Shi}$^{1},$%
{\small Xiwen Zhu}$^{1}$ {\small and Kelin Gao}$^{1}$}
\address{{\small \ }$^{1}$Laboratory of Magnetic Resonance and Atomic and Molecular
Physics,Wuhan Institute of Physics and Mathematics,The Chinese Academy of
Sciences,Wuhan,430071, P.R.China. \\
$^{2}$Department of Physics,Hunan Normal
University,Changsha,410081,P.R.China.\\
$^{3}$Department of Physics,University of
Waterloo,Waterloo,Ontario,N2L3G1,Canada.\\
$^{4}$Department of Physics and Atmospheric Science,Drexel
University,Philadelphia, PA19104,USA.}
\date{\today}
\maketitle

\begin{abstract}
Based upon our previous works ( Eur.Phys.J.D 6, 319(1999); Chin.Phys.Lett.
18, 236(2001)), we develop a classical approach to calculate the high-order
harmonic generation of the laser driven atoms and molecules. The Coulomb
singularities in the system have been removed by a regularization procedure.
Action-angle variables have been used to generate the initial microcanonical
distribution which satisfies the inversion symmetry of the system. The
numerical simulation show, within a proper laser intensity, a harmonic
plateau with only odd harmonics appears. At higher intensities, the spectra
become noisier because of the existence of chaos. With further increase in
laser intensity, ionization takes place, and the high-order harmonics
disappear. Thus chaos introduces noise in the spectra, and ionization
suppresses the harmonic generation, with the onset of the ionization follows
the onset of chaos.
\end{abstract}



\section{Introduction}

The development of the high-power femtosecond laser has stimulated the
investigation of the multi-photon processes of atoms and molecules
interacting with intense laser fields [1-9]. Recently, there are many
theoretical and experimental references about these multiphoton process.
Within a proper intensity region, lots of odd harmonics of laser are
generated by atomic and molecular gases [1-5,14]. The harmonic structure
distributes as a plateau which is cut off at a special high-order harmonic.
When the intensity increases, the ionization channel is opened and the
high-order harmonics disappear. The above threshold ionization (ATI) occurs
when the laser intensity is sufficient power (above $10^{13}W/cm^{2}$).
Various structures (plateau, angular distribution, etc.) in ATI spectra have
been detailed in Ref.[6-8]. The above threshold dissociation (ATD) and
dissociation-ionization of molecular systems are also reported [9-10].
Studying the laser-matter interaction deeply, not only can obtain new
knowledge of the interacting mechanism, but also can provide widely
application in the generation of high-order coherent harmonics, $X$-rays
laser and $\gamma $-rays laser.

The classical dynamics of most laser-driven systems is generally chaotic,
due to the existence of nonlinearity. Chaos usually manifests itself as some
control parameters (initial energy, laser intensity, laser frequency, etc.)
are varied. The microscopic systems, in particular those involving atoms and
molecules, are governed by a Hamiltonian. To study these systems are of
great importance in the context of quantum-classical correspondence. At the
investigated high power laser intensity $(10^{13}\symbol{126}10^{15}W/cm^{2})
$, the electric field of the laser is equal in strength to the Coulomb field
of the nuclei [8], and the laser field can not be looked as a perturbation
to the field-free system, for the states themselves are no longer
independent of the laser. Generally, the microscopic systems (atoms or
molecules) are intrinsically quantum mechanical systems, thus they must be
described by quantum mechanics. However, due to the presence of intense
laser, the exact calculation even numerical simulation based upon quantum
mechanics tend to be very difficult to perform. Fortunately, classical
approach to these similar problems is useful for providing physical insight
into dynamics processes [10-13]. Classical chaos associating with the
microwave ionization of atomic hydrogen reveals that the detailed mechanism
of atomic ionization in terms of transport in phase space [11-12]. Classical
prediction for scaled frequencies has been verified by quantum calculations
[12-13], and, in turn, has confirmed the dynamical significance of classical
chaos. The harmonic generation (HG) of laser driven hydrogen atoms is
simulated with classical Monte-Carlo method [14], the numerical results are
qualitatively consistent with the quantum mechanic results and the
experimental observation. For the molecular systems, the classical
calculation is a good first step, since length and energy scales are often
large enough for classical mechanics to be at least approximately valid.
Using softened potential model, the classical dynamics of the
one-dimensional hydrogen molecular ion $H_{2}^{+}$ interacting with an
intense laser pulse are detailed [15-16].

There are two important and realistic examples of laser driven systems,
where nonlinear dynamics has played a major role, which have been
traditionally studied. One is the laser driven hydrogen atom that is the
fundamental system in atomic physics, the other is the laser driven hydrogen
molecular ion $H_{2}^{+}$, which is the fundamental diatomic molecular
system in molecular physics. Bellow, we shall show our classical calculation
of the high-order harmonic generation of these two fundamental systems. The
outline of this paper is as follows. The regularized model and the
corresponding initial microcanonical distribution are presented in the next
section. In section III, we show the numerical results in details. A briefly
summary and discussion is given out in the last section.

\section{Model and initial microcanonical distribution}

\subsection{Regularized model of laser driven hydrogen atom}

In this article, we consider a classical hydrogen atom, with an infinite
mass nucleus fixed at the origin of coordinates, interacting with a high
intense laser field which is linearly polarized along the $z-axis$, and with
the electric field component $\varepsilon (t)$. Thus the Hamiltonian in
atomic units in Cartesian coordinates is 
\begin{eqnarray}
H &=&H_{0}+H_{i,}  \nonumber \\
H_{0} &=&\frac{1}{2}p^{2}-\frac{1}{r},H_{i}=-z\varepsilon (t).
\end{eqnarray}
where, $r=\sqrt{x^{2}+y^{2}+z^{2}}$ and $p^{2}=p_{x}^{2}+p_{y}^{2}+p_{z}^{2}$%
. Apparently, the above Hamiltonian has a Coulomb singularity corresponding
to electron-nucleus collision. To remedy the singularity, we introduce the
parabolic coordinates $(u,v,\phi )$%
\begin{equation}
x=uv\cos \phi ,y=uv\sin \phi ,z=(v^{2}-u^{2})/2.
\end{equation}
and a new fictive time scale $\tau $%
\begin{equation}
dt/d\tau =G(u,v)=v^{2}+u^{2}.
\end{equation}
In order to implement regularization, following the notation of Szebenhely
[17], regarding the motion time $t$ and the negative total energy $-E$ as
generalized coordinate and generalized momentum respectively, then the
Hamiltonian function in extended space becomes into 
\begin{equation}
H^{*}=H-E\equiv 0=H^{*}(x,y,z,t;p_{x},p_{y},p_{z},-E)
\end{equation}
For the sake of obtaining the regularized Hamiltonian, we introduce the
third category generating function $F_{3}(p_{x},p_{y},p_{z},-E;u,v,\phi ,t)$%
, then obtain 
\begin{equation}
x(u,v,\phi )=-\frac{\partial F_{3}}{\partial p_{x}},y(u,v,\phi )=-\frac{%
\partial F_{3}}{\partial p_{y}},z(u,v,\phi )=-\frac{\partial F_{3}}{\partial
p_{z}},t(u,v,\phi )=-\frac{\partial F_{3}}{\partial E}.
\end{equation}
So the third category generating function $F_{3}(p_{x},p_{y},p_{z},-E;u,v,%
\phi ,t)$ is in the form of 
\begin{eqnarray}
F_{3} &=&-xp_{x}-yp_{y}-zp_{z}+Et,  \nonumber \\
&=&-uv\cos \phi p_{x}-uv\sin \phi p_{y}-\frac{1}{2}(v^{2}-u^{2})p_{z}+Et.
\end{eqnarray}
The new momenta $(p_{u},p_{v},p_{\phi },-E)$ are related with old momenta $%
(p_{x},p_{y},p_{z},-E)$ by 
\begin{eqnarray}
p_{u} &=&-\frac{\partial F_{3}}{\partial u}=v\cos \phi p_{x}+v\sin \phi
p_{y}-up_{z},  \nonumber \\
p_{v} &=&-\frac{\partial F_{3}}{\partial v}=u\cos \phi p_{x}+u\sin \phi
p_{y}-vp_{z},  \nonumber \\
p_{\phi } &=&-\frac{\partial F_{3}}{\partial \phi }=-uv\sin \phi
p_{x}+uv\cos \phi p_{y},  \nonumber \\
E &=&\frac{\partial F_{3}}{\partial t}=E.
\end{eqnarray}
Then the regularized Hamiltonian can be expressed as 
\begin{eqnarray}
K &=&\frac{dt}{d\tau }H^{*}=G(u,v)(H-E)\equiv 0,  \nonumber \\
&=&\frac{1}{2}[p_{u}^{2}+p_{v}^{2}+(u^{-2}+v^{-2})p_{\phi
}^{2}]-2-E(u^{2}+v^{2})-\frac{1}{2}(v^{4}-u^{4})\varepsilon (t).
\end{eqnarray}
If we define 
\begin{eqnarray}
K_{u} &=&\frac{1}{2}[p_{u}^{2}+u^{-2}p_{\phi }^{2}]-1-Eu^{2}+\frac{1}{2}%
u^{4}\varepsilon (t),  \nonumber \\
K_{v} &=&\frac{1}{2}[p_{v}^{2}+v^{-2}p_{\phi }^{2}]-1-Ev^{2}-\frac{1}{2}%
v^{4}\varepsilon (t).
\end{eqnarray}
In the field-free case, $\varepsilon (t)=0$, $K_{u}(=-K_{v})$ is the
component of the Laplace-Runge-Lenz vector along $z-axis$, i.e., $%
K_{u}=-K_{v}=R_{z}$. The equations of motion can be derived from the above
regularized Hamiltonian, as the following: 
\begin{equation}
\frac{d\phi }{d\tau }=\frac{\partial K}{\partial p_{\phi }}=p_{\phi
}(u^{-2}+v^{-2}),\text{ }\frac{dp_{\phi }}{d\tau }=-\frac{\partial K}{%
\partial \phi }=0,
\end{equation}
\begin{equation}
\frac{du}{d\tau }=\frac{\partial K}{\partial p_{u}}=p_{u},\text{ }\frac{%
dp_{u}}{d\tau }=-\frac{\partial K}{\partial u}=2Eu+p_{\phi
}^{2}u^{-3}-2u^{3}\varepsilon (t),
\end{equation}
\begin{equation}
\frac{dv}{d\tau }=\frac{\partial K}{\partial p_{v}}=p_{v},\text{ }\frac{%
dp_{v}}{d\tau }=-\frac{\partial K}{\partial v}=2Ev+p_{\phi
}^{2}v^{-3}+2v^{3}\varepsilon (t),
\end{equation}
and 
\begin{equation}
\frac{dE}{d\tau }=\frac{\partial K}{\partial t}=\frac{1}{2}(u^{4}-v^{4})%
\frac{d\varepsilon (t)}{dt},\text{ }\frac{dt}{d\tau }=-\frac{\partial K}{%
\partial E}=G(u,v).
\end{equation}
Apparently, $p_{\phi }$ is a constant which corresponding to the component
of angular momentum along $z-axis$.

\subsection{Regularized model of laser driven hydrogen molecular ion $%
H_{2}^{+}$}

Within Born-Oppenheimer approximation, we can assume that two protons are
fixed at the positions $A$ and $B$, with a distance $R$ away from each
other, in the locations $(0,0,-R/2)$ and $(0,0,R/2)$ of the Cartesian
coordinates system. A single electron at $(x,y,z)$ is subject to the Coulomb
attraction of both protons and the interaction of the laser. Let $r_{A\text{ 
}}$ be its distance from $A$, and $r_{B\text{ }}$ be its distance from $B$.
Within the dipole approximation, the Hamiltonian in atomic units is 
\begin{eqnarray}
H &=&H_{0}+H_{i,}  \nonumber \\
H_{0} &=&\frac{1}{2}p^{2}-1/r_{A}-1/r_{B},H_{i}=-z\varepsilon (t).
\end{eqnarray}
where, $\varepsilon (t)$ is the electric field of the laser pulse, $H_{i}$
is the interacting Hamiltonian. Apparently, the above Hamiltonian has
singularities at points $r_{A\text{ }}=0$ and $r_{B}=0$, which corresponds
to electron-proton collision. To overcome this barrier, regularization has
to be performed. Define the new coordinates $(u,v,\phi )$ as 
\begin{equation}
x=-\frac{R}{2}\sin u\sinh v\cos \phi ,y=-\frac{R}{2}\sin u\sinh v\sin \phi
,z=\frac{R}{2}\cos u\cosh v.
\end{equation}
and introduce the new fictive time scale $\tau $ satisfying 
\begin{equation}
dt/d\tau =G(u,v)=r_{A}r_{B}=R^{2}(\cosh ^{2}v-\cos ^{2}u)/4.
\end{equation}
Similar to the previous subsection, regarding the motion time $t$ and the
negative total energy $-E$ as generalized coordinate and generalized
momentum respectively, then we can write the Hamiltonian function in
extended space as 
\begin{equation}
H^{*}=H-E\equiv 0=H^{*}(x,y,z,t;p_{x},p_{y},p_{z},-E)
\end{equation}
Introducing the third category generating function $%
F_{3}(p_{x},p_{y},p_{z},-E;u,v,\phi ,t)$, which satisfies 
\begin{equation}
F_{3}=-\frac{R}{2}\sin u\sinh v\cos \phi p_{x}-\frac{R}{2}\sin u\sinh v\sin
\phi p_{y}-\frac{R}{2}\cos u\cosh vp_{z}+Et.
\end{equation}
Thus momenta $(p_{u},p_{v},p_{\phi },-E)$ are related with old momenta $%
(p_{x},p_{y},p_{z},-E)$ by 
\begin{eqnarray}
p_{u} &=&-\frac{\partial F_{3}}{\partial u}=\frac{R}{2}\cos u\sinh v\cos
\phi p_{x}+\frac{R}{2}\cos u\sinh v\sin \phi p_{y}-\frac{R}{2}\sin u\sinh
vp_{z},  \nonumber \\
p_{v} &=&-\frac{\partial F_{3}}{\partial v}=\frac{R}{2}\sin u\cosh v\cos
\phi p_{x}+\frac{R}{2}\sin u\cosh v\sin \phi p_{y}-\frac{R}{2}\sin u\cosh
vp_{z},  \nonumber \\
p_{\phi } &=&-\frac{\partial F_{3}}{\partial \phi }=-\frac{R}{2}\sin u\sinh
v\sin \phi p_{x}+\frac{R}{2}\sin u\sinh v\cos \phi p_{y},  \nonumber \\
E &=&\frac{\partial F_{3}}{\partial t}=E.
\end{eqnarray}
And the regularized Hamiltonian is 
\begin{eqnarray}
K &=&\frac{dt}{d\tau }H^{*}=G(u,v)(H-E)\equiv 0,  \nonumber \\
&=&\frac{1}{2}[p_{u}^{2}+p_{v}^{2}+(\csc ^{2}u+\csc h^{2}v)p_{\phi
}^{2}]-R\cosh v-\frac{(E-H_{i})}{4}R^{2}(\cosh ^{2}v-\cos ^{2}u).
\end{eqnarray}
where, $H_{i}=-z\varepsilon (t)=(-R/2)\cos u\cosh v\varepsilon (t)$.
Obviously, $p_{\phi }$ is a constant, which corresponds to the component of
the angular momentum which along the $z-axis$. If $p_{\phi }$ is equal to
zero, the electron is constrained in a plane which can be chosen as $y=0$,
corresponding to a two-dimensional motion. In the field-free case, $%
\varepsilon (t)=0$, this two-dimensional model corresponds to the classical
hydrogen molecular ion in ground-state. Equations of the motion for the
ground-state hydrogen molecular ion interacting with laser field can be
easily obtained from the regularized Hamiltonian, i.e., 
\begin{eqnarray}
\frac{du}{d\tau } &=&\frac{\partial K}{\partial p_{u}}=p_{u},  \nonumber \\
\frac{dp_{u}}{d\tau } &=&-\frac{\partial K}{\partial u}=\frac{ER^{2}}{4}\sin
(2u)+\varepsilon (t)(z\frac{\partial G}{\partial u}+G\frac{\partial z}{%
\partial u}),
\end{eqnarray}
\begin{eqnarray}
\frac{dv}{d\tau } &=&\frac{\partial K}{\partial p_{v}}=p_{v},  \nonumber \\
\frac{dp_{v}}{d\tau } &=&-\frac{\partial K}{\partial v}=R\sinh v+\frac{ER^{2}%
}{4}\sinh (2v)+\varepsilon (t)(z\frac{\partial G}{\partial v}+G\frac{%
\partial z}{\partial v}),
\end{eqnarray}
and 
\begin{equation}
\frac{dE}{d\tau }=\frac{\partial K}{\partial t}=zG\frac{d\varepsilon (t)}{dt}%
,\text{ }\frac{dt}{d\tau }=-\frac{\partial K}{\partial E}=G(u,v).
\end{equation}
In the field-free case, defining 
\begin{eqnarray}
K_{u} &=&\frac{1}{2}(p_{u}^{2}+p_{\phi }^{2}\csc ^{2}u)+\frac{ER^{2}}{4}\cos
^{2}u,  \nonumber \\
K_{v} &=&\frac{1}{2}(p_{v}^{2}+p_{\phi }^{2}\csc h^{2}v)-R\cosh v-\frac{%
ER^{2}}{4}\cosh ^{2}v.
\end{eqnarray}
thus $K_{u}(=-K_{v})$ is a constant of the field-free motion, they are
related to $\gamma $ and $\Omega $ by 
\begin{equation}
K_{u}=-K_{v}=-\gamma R^{2}/4=ER^{2}/4+\Omega /2.
\end{equation}
Constants $\Omega $ and $\gamma $ are first introduced by Erickson [18] and
Strand [19] respectively, which have the following forms 
\begin{eqnarray}
\gamma &=&-E-2\Omega /(mR^{2}),  \nonumber \\
\Omega &=&\overrightarrow{L_{A}}\cdot \overrightarrow{L_{B}}+emR^{2}(\cos
\theta _{A}-\cos \theta _{B}).
\end{eqnarray}
Here, $\overrightarrow{L_{A}}$ and $\overrightarrow{L_{B}}$ are the angular
momentum vector of the motion around nucleus $A$ and $B$ respectively, $%
\theta _{A}$ and $\theta _{B}$ are the angle from the vector $%
\overrightarrow{r_{A}}$ and $\overrightarrow{r_{B}}$ to positive $z-axis$
respectively, $m$ and $e$ are the mass and the charge of the electron
respectively.

\subsection{Action-angle variables and initial distributions}

The action-angle variables for separated systems are defined as 
\begin{equation}
I_{i}=\frac{1}{2\pi }\oint p_{i}dq_{i},\text{ }\theta _{i}=\frac{\partial }{%
\partial I_{i}}\int p_{i}dq_{i}=\int \frac{\partial p_{i}}{\partial H}\frac{%
\partial H}{\partial I_{i}}dq_{i}.
\end{equation}
The motion of the classical field-free hydrogen atom is periodic, then it
need only a pair of conjugated action-angle variables [20] 
\begin{equation}
H_{0}=E_{0}=-\frac{1}{2I^{2}}.
\end{equation}
The corresponding angle is given by the Kepler's equation 
\begin{equation}
\theta =u-e\sin u.
\end{equation}
This means $\theta $ is the mean anomaly of the free orbits. The eccentric
anomaly $u$ can be obtained from $r=a(1-e\cos u)$. Here, $r$ is the distance
from the electron to the origin, $a$ is the instantaneous semimajor axis
which satisfies $a=-1/(2E_{0})$, $e$ is the eccentricity of the free orbits
satisfying $e=\sqrt{2E_{0}L^{2}+1}$, $L$ is the total angular momentum.

The motion of the ground-state hydrogen molecular ion $H_{2}^{+}$ is
quasi-periodic, it need two pairs of action-angle variables. With elliptic
integrals, action $I_{u}$ can be expressed as follows. 
\begin{equation}
I_{u}=\left\{ 
\begin{array}{ll}
\frac{\sqrt{8K_{u}-2E_{0}R^{2}}}{\pi }F_{1}(\frac{\pi }{2},\sqrt{1+\frac{%
4K_{u}}{E_{0}R^{2}-4K_{u}}}), & for\text{ }K_{u}>0, \\ 
\sqrt{-E_{0}R^{2}/\pi }, & for\text{ }K_{u}=0, \\ 
\frac{\sqrt{-2E_{0}R^{2}}}{\pi }[F_{1}(\frac{\pi }{2},\sqrt{1-\frac{4K_{u}}{%
E_{0}R^{2}}})-\frac{4K_{u}}{E_{0}R^{2}}F_{2}(\frac{\pi }{2},\sqrt{1-\frac{%
4K_{u}}{E_{0}R^{2}}})], & for\text{ }K_{u}<0.
\end{array}
\right.
\end{equation}
where, the first category elliptic integral $F_{1}(\varphi ,k)$ and the
second category elliptic integral $F_{2}(\varphi ,k)$ are in the form of 
\begin{eqnarray*}
F_{1}(\varphi ,k) &=&\int\limits_{0}^{\varphi }\sqrt{1-k^{2}\sin ^{2}x}dx, \\
F_{2}(\varphi ,k) &=&\int\limits_{0}^{\varphi }1/\sqrt{1-k^{2}\sin ^{2}x}dx.
\end{eqnarray*}

Generally, a single trajectory lacks the inversion symmetry of the real
physical systems. So the spectrum obtained from a single trajectory exhibits
unphysical even harmonics. A natural way to remedy the unphysical even
harmonics is to consider an ensemble of trajectories, evolving from an
initial microcanonical distribution with inversion symmetry. For a chaotic
system, the initial distribution can be generated with Monte-Carlo method.
However, for an integrable system, the distribution generated by Monte-Carlo
method does not possess of ergodicity. We find that the points on the same
equienergy surface generated by action-angle variables with regular steps
possess of good ergodicity. To reconstruct the inversion symmetry, there
must exist pairs of $(x_{0},y_{0},z_{0};p_{x_{0}},p_{y_{0}},p_{z_{0}})$ and $%
(-x_{0},-y_{0},-z_{0};-p_{x_{0}},-p_{y_{0}},-p_{z_{0}})$ in the initial
distribution. For the regularized model for hydrogen atom, it corresponds to
pairs of $(u_{0},v_{0},\phi _{0};p_{u_{0}},p_{v_{0}},p_{\phi _{0}})$ and $%
(u_{0},v_{0},\phi _{0}+\pi ;p_{u_{0}},p_{v_{0}},p_{\phi _{0}})$. And for the
regularized hydrogen molecular ion $H_{2}^{+}$, it corresponds to pairs of $%
(u_{0},v_{0},\phi _{0};p_{u_{0}},p_{v_{0}},p_{\phi _{0}})$ and $(u_{0}+\pi
,v_{0},\phi _{0};p_{u_{0}},p_{v_{0}},p_{\phi _{0}})$.

\section{Numerical simulation}

The numerical computational procedure is based upon the classical trajectory
Monte Carlo (CTMC) method [20-21]. CTMC simulation procedure involves three
stages, (i) choice of initial conditions, (ii) numerical integration of
equation of motion, and (iii) categorization of each trajectory as
excitation, charge transfer or ionization. In the process of numerical
integrating, numerical accuracy and computing time are two primary aspects
that must be considered. We use the fourth-order Runge-Kutta method with
variable steps to perform the numerical calculation. Note also that computer
can not deal with singularity, which corresponds to electron-nucleus
collision, to overcome this difficulty, we have implemented regularization.
With complete regularization, numerical simulation can be established with
required precision before, at, and, after collision successfully.

To obtain the harmonic spectra, our procedure also calculate the averaged
dipole moment of the excited trajectories with pairs of inversion symmetric
initial conditions in the same distribution. Having determined the actual
trajectories of the electron, one can easily obtain the component of the
averaged dipole moment, which along the laser polarization direction. Then
the harmonic spectra of the driven dipole is straightforwardly obtained from
it's power spectra 
\begin{equation}
D(\omega )=\lim_{t\rightarrow +\infty }\frac{1}{t}\left|
\int\limits_{0}^{t}\left\langle \mu (\tau )\right\rangle e^{i\omega \tau
}d\tau \right| ^{2}.
\end{equation}
where, $\left\langle \mu (\tau )\right\rangle $ is the component of the
averaged dipole moment along the laser polarization direction. In our
calculation, the electric fields of the ultrashort laser pulses are chosen
as 
\begin{equation}
\varepsilon (t)=\left\{ 
\begin{array}{ll}
E_{M}\sin ^{2}(\omega _{L}t/40)\sin (\omega _{L}t), & ~~~~for\text{ }0\leq
t\leq 40\pi /\omega _{L}, \\ 
0, & ~~~~otherwise.{\rm ~~~~~~~~~~~~~~~~~~~~~~~~}
\end{array}
\right.
\end{equation}
The maximum electric field strength $E_{M}$ is related to the laser
intensity $I=\sqrt{\varepsilon _{0}/\mu _{0}}E_{M}^{2}/2$, and $\omega _{L}$
is the angular frequency, and period $T_{L}$ is equal to $2\pi /\omega _{L}$.

The time-dependent dipole moment $\mu _{z}(t)$ of a single trajectory along
the laser polarization direction sensitively depends on the laser intensity.
Fig.1 shows the time evolution of the dipole of the hydrogen molecular ion
with initial energy $E_{0}=-1.1034$ $hatree$, internuclear distance $R=2.00$ 
$bohr$, laser wavelength $\lambda =600$ $nm$ and different laser intensity.
And some dipoles of the hydrogen atom and their power spectra are presented
in Fig.2, with initial energy $E_{0}=-0.5$ $hatree$, laser wavelength $%
\lambda =532$ $nm$ and different laser intensity. With the increasing of the
laser intensity, the oscillations of the dipole moments are modulated
gradually, and it follows a regular pattern with both high and low frequency
components. Such patterns are independent of initial conditions. When the
laser intensity is large enough, ionization takes place, which strongly
modifies the subsequent time evolution of the dipole moment.

The power spectra obtained from a single trajectory of the laser driven
hydrogen molecular ion, which with initial energy $E_{0}=-1.1034$ $hatree$,
internuclear distance $R=2.00$ $bohr$, laser wavelength $\lambda =600$ $nm$
and different laser intensity, are presented in Fig.3. The first one
corresponds to the field-free case. For the hydrogen atom, it possesses of
only regularly decreasing peaks locating at multiples (harmonics) of
Kepler's frequency, that is, $\omega _{n}=n\omega _{0}$, for $n=1,2,3\cdot
\cdot \cdot $, which manifests that the free motion is periodic. While for
the hydrogen molecular ion, the free motion is quasi-periodic, it appears
regularly decreasing peaks locating at two different characteristic
frequencies and their combinations, i.e., at $n\omega _{01}+m\omega _{02}$,
for $n=0,1,2,3\cdot \cdot \cdot $, $m=0,1,2,3\cdot \cdot \cdot $ and $n+m>0$%
. For the laser driven systems, the peaks of harmonic spectra depend
strongly on the laser intensity. In addition to the origin peaks, a dominant
line at the laser frequency $\omega _{L}$ appears, which is the Rayleigh
component in the light scattered by the atoms and molecular ions. High-order
harmonics, which consist of both odd and even components, do appear in the
spectra, and their orders and strengths increase with the increasing of the
laser intensity. For a low laser intensity, i.e., in the perturbative
regime, the characteristic peaks dominate the power spectrum. For a proper
laser intensity, the plateau structure containing both odd and even
harmonics appears. For a larger laser intensity, because of the presence of
chaos, the spectra become noisier even the motion is bound. For a strong
enough laser intensity, the motion becomes unbound and the corresponding
spectrum is dominated by a very noise background, which is generated by the
ionized electrons, the only line surviving being the one at the laser
frequency, which corresponds to the light scattered by the asymptotically
free electron Thomson scattering.

To eliminate the unphysical even harmonics, averaging an ensemble of
trajectories evolving from an inversion symmetric distribution is necessary.
The time evolution of the averaged dipole, which evolve from a
microcanonical ensemble of 5000 trajectories, are presented in Fig.4, the
first one corresponds to the hydrogen atom with initial energy $E_{0}=-0.5$ $%
hatree$, laser wavelength $\lambda =532$ $nm$ and laser intensity $%
I=5.0\times 10^{14}$ $W/cm^{2}$, the second one corresponds to the hydrogen
molecular ion with initial energy $E_{0}=-1.1034$ $hatree$, internuclear
distance $R=2.00$ $bohr$, laser wavelength $\lambda =600$ $nm$ and laser
intensity $I=1.0\times 10^{14}$ $W/cm^{2}$. Comparing with the evolution of
the dipole of a single trajectory, one can easily find that the oscillation
of the averaged dipole is smooth and it globally follows the laser
oscillation.

The spectra obtained from an ensemble of trajectories are showed in Fig.5
and Fig.6. Fig.5 is the harmonic spectra of the ground-state hydrogen atoms
interacting with the laser pulses with $\lambda =532$ $nm$ and different
laser intensity. Fig.6 is harmonic spectra of the hydrogen molecular ion
with initial energy $E_{0}=-1.1034$ $hatree$, internuclear distance $R=2.00$ 
$bohr$, laser wavelength $\lambda =600$ $nm$ and different laser intensity.
As a consequence of averaging process, the unphysical even harmonics are
remedied really. Within a proper laser intensity range, the plateau
structure that only possesses of odd harmonics appears. As pointed out
previously, at a higher intensity, the spectra become noisier even the
ionization does not happen because of the effects of chaos. This indicates
that the chaos cause the noise of the harmonic spectra. When the laser
intensity is high enough, the ionization takes place, thus the noise
background conceals the high-order harmonics. This means that the onset of
ionization follows the onset of chaos and the ionization suppresses the
harmonic generation.

\section{Summary and discussion}

In summary, within the Born-Oppenheimer approximation and using the
classical trajectory method,we have calculated the high-order harmonic
generation spectra of the hydrogen atom and the hydrogen molecular ion
interacting with ultrashort intense laser pulses. The other multi-photon
phenomena, such as multi-photon ionization and above threshold dissociation,
can also be simulated. In our subsequent calculations, the dynamics of the
electron is investigated by numerical integrating the equations of motion
using regularized coordinates. To eliminate the unphysical even harmonics of
a single trajectory, averaging over an ensemble of trajectories evolving
from an initial microcanonical distribution with inversion symmetry is
necessary. Such distribution is constructed using action-angle variables. A
plateau structure in the spectra with only odd harmonics is observed within
a proper laser intensity range of about $10^{14}$ $W/cm^{2}$. From our
numerical results, we observe that the high order harmonics are cut off at a
special order harmonic. At higher laser intensities, chaos introduces noise
into the spectra even though the motion is still bound. Finally as the
intensity is further increased, ionization takes place, and the harmonics
disappear.

These results are qualitatively consistent with recent quantum calculations
[16] and experimental observations [1-5], but the cutoff order $N_{m}$ of
the plateau structure is not precisely consistent with formula $%
N_{m}=(I_{p}+3.17U_{p})/\hbar \omega _{_{L}}$[27], here, $I_{p}$ is the
ionization potential and $U_{p}=e^{2}E_{M}^{2}/4m_{e}\omega _{_{L}}^{2}$
denotes the quiver energy or the ponderomotive energy of an electron. As an
example, when the laser intensity $I=10^{14}$ $W/cm^{2}$ and wavelength $%
\lambda =600$ $nm$, the ionization potential $I_{p}$ of the ground-state
hydrogen molecular ion is $1.1034$ $hatree$ $(29.77eV=14.50$ $\hbar \omega
_{_{L}})$, the quiver energy $U_{p}=3.36eV$ $(=1.63$ $\hbar \omega _{_{L}})$%
, then $N_{m}=19$, and when $I=7.5\times 10^{13}$ $W/cm^{2}$, $N_{m}=18$;
however, from our simulation, the harmonic plateau are both cut off at $17$,
and for very large intensity (above $10^{15}$ $W/cm^{2}$) the high-order
harmonics are concealed by the noises. To obtain quantitative results, we
have to integrate the time-dependent Schr$\stackrel{..}{o}$dinger equation,
this can be realized with the split-operator method [22]. For the hydrogen
molecular ion, within BOA, a proper internuclear distance $R$ will enhance
the high-order harmonic generation[16], and it will be interesting to go
beyond the Born-Oppenheimer approximation to investigate what further
interesting insights can be obtained when the nuclear motion is taken into
account [23-24].

In our model, we only consider the non-relativistic case with dipole
approximation. When the laser is sufficiently intense, photoelectrons of
relativistic energies can be produced, necessitating a fully relativistic
treatment [25]. The dipole approximation is no longer valid, and the
magnetic field is not only present, but acquires an importance similar to
that of the electric field. And even before the appearance of the
relativistic photoelectrons, the effects of the magnetic field may be very
important too [26]. Due to the wiggly motion and the acceleration of the
electron near the outermost turning points induced by the magnetic field,
the cut-off order of the harmonic plateau maybe higher.

\begin{center}
{\large Acknowledgment}
\end{center}

The work is supported by the National Natural Science Foundations of China
under Grant No. 19874019, 19904013 and 1990414. The author Chaohong Lee
thanks very much for the help of Dr. Haoseng Zeng, Dr. Ming He and Dr.
Zongxiu Nie. 
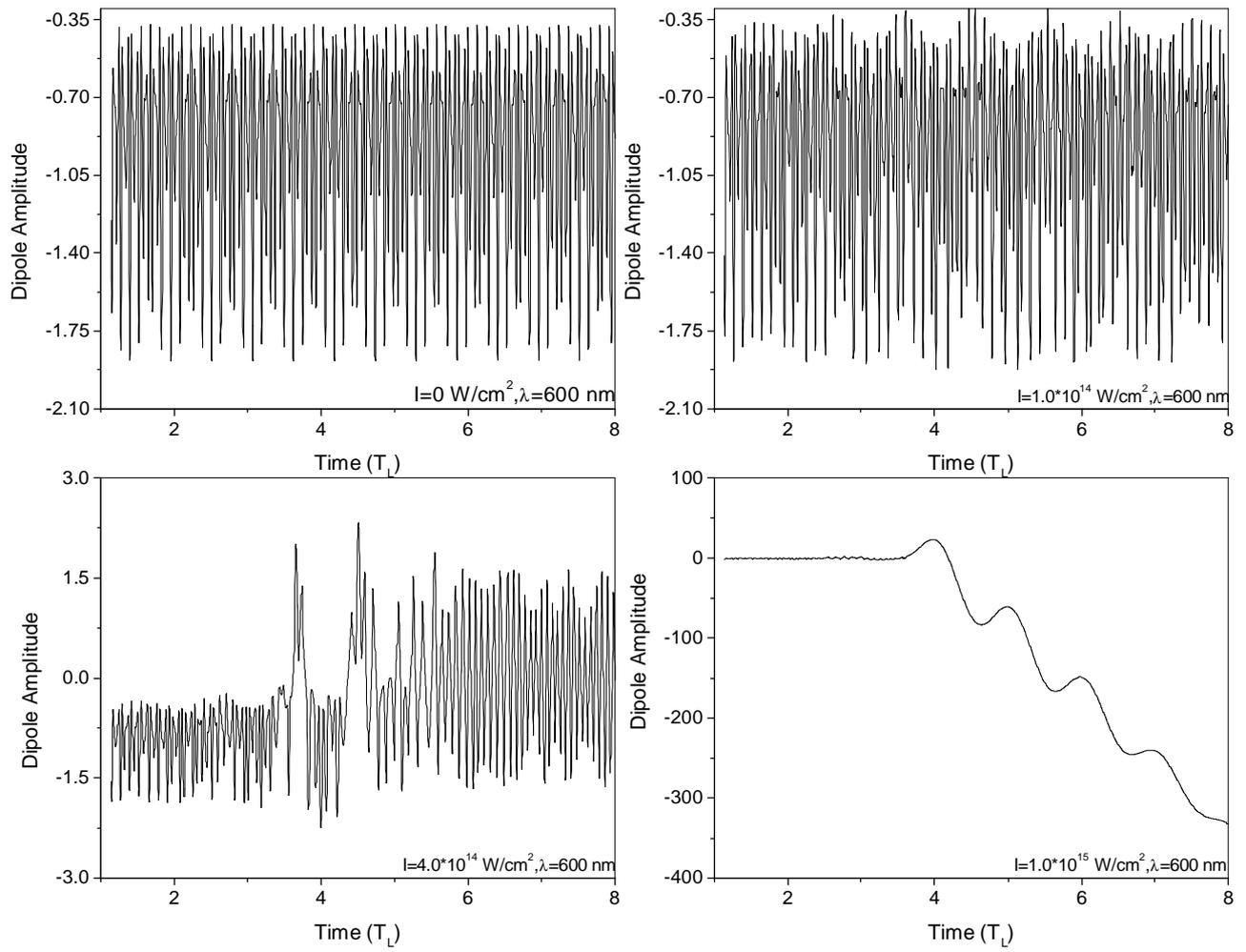
\begin{figure}[tbp]
\caption{Temporal variation of the dipole moment of a single electronic
trajectory of the hydrogen molecular ion for different laser parameters with
initial energy $E_{0}=-1.1034$ $hatree$ and internuclear distance $R=2.00$ $%
bohr$. }
\end{figure}
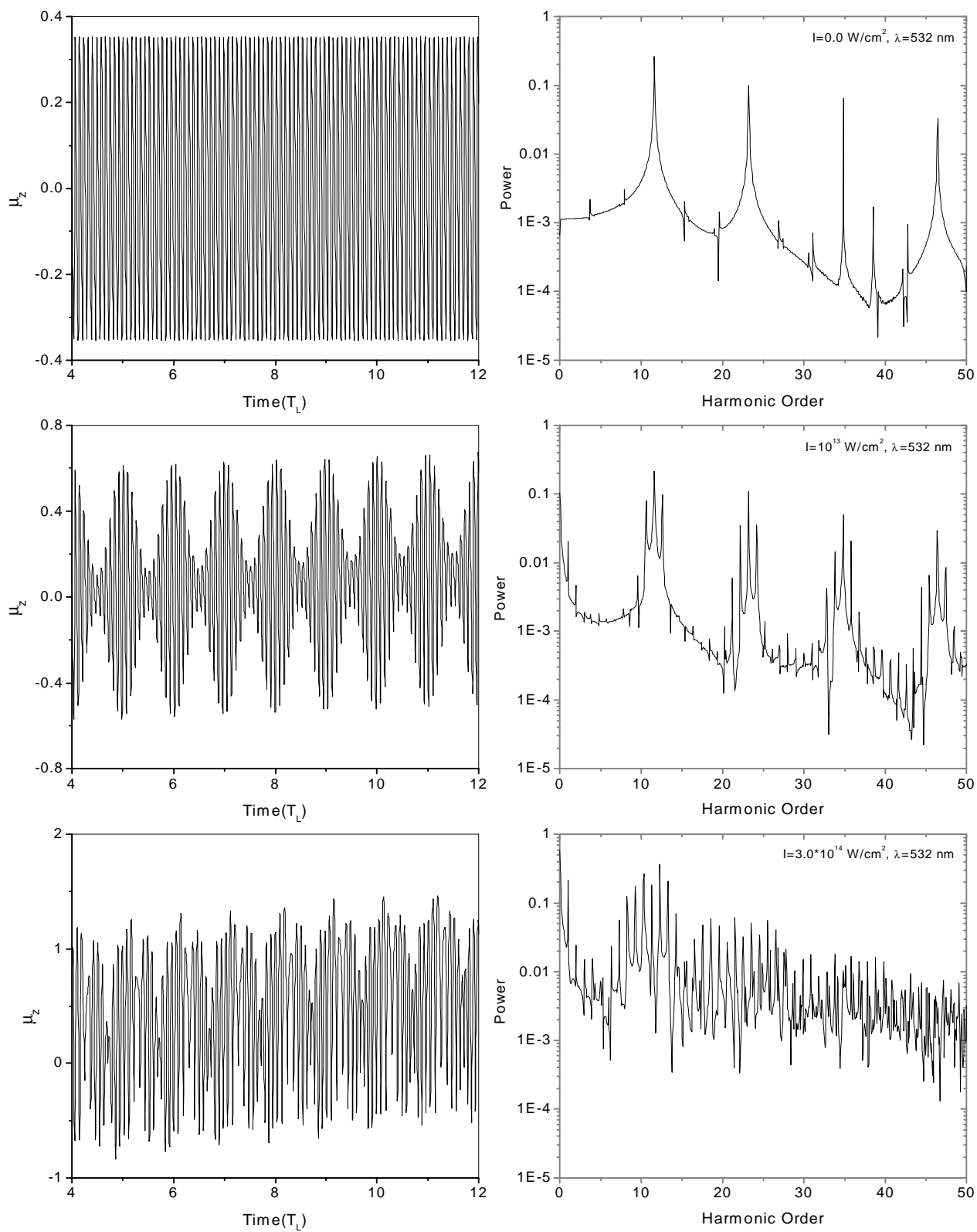
\begin{figure}[tbp]
\caption{The dipole and their power spectra of a single electronic
trajectory of the hydrogen atom for different laser parameters with initial
energy $E_{0}=-0.5$ $hatree$. }
\end{figure}
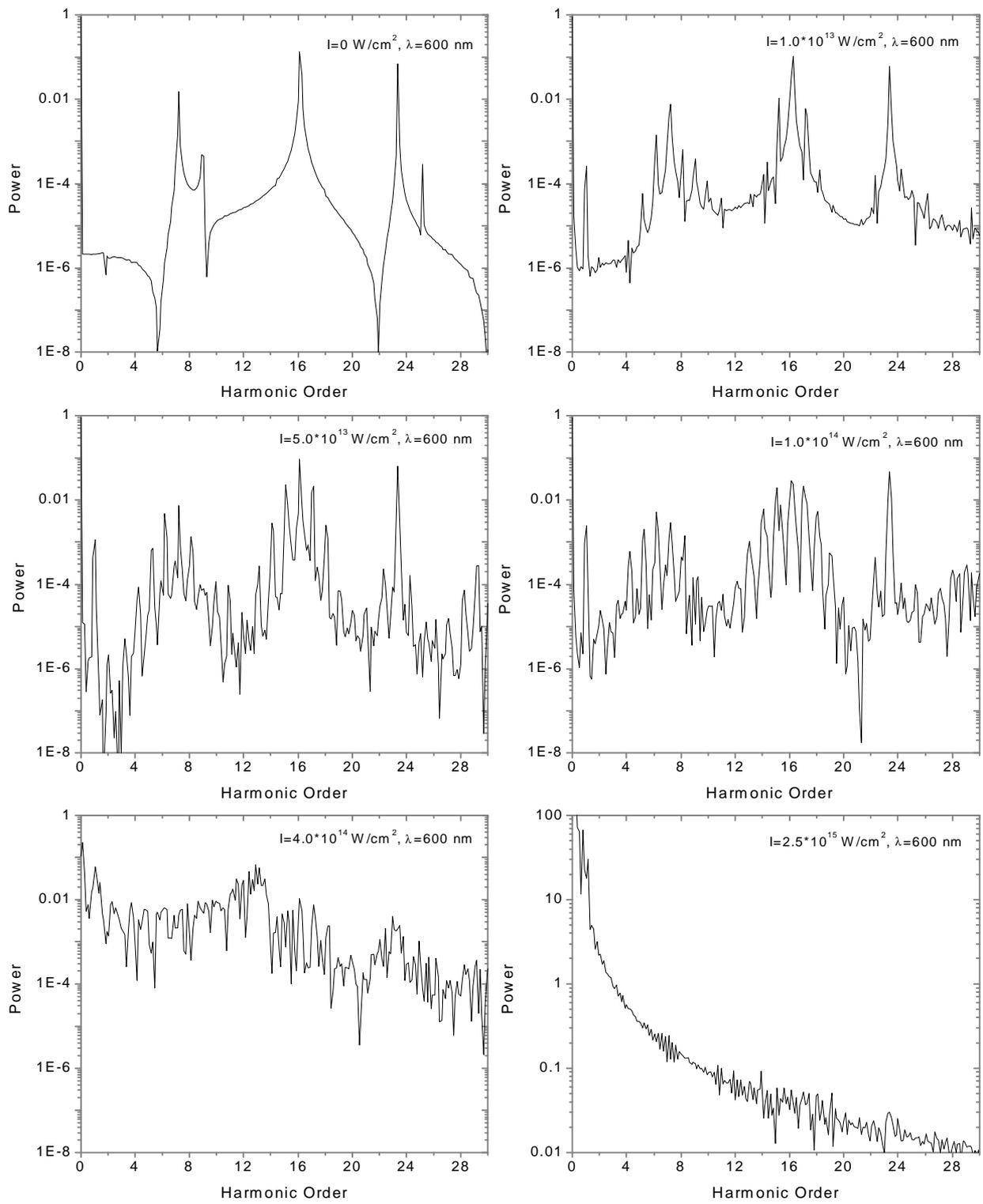
\begin{figure}[tbp]
\caption{Power spectra of the dipole of a single electronic trajectory of
the hydrogen molecular ion for different laser parameters with initial
energy $E_{0}=-1.1034$ $hatree$ and internuclear distance $R=2.00$ $bohr$. }
\end{figure}
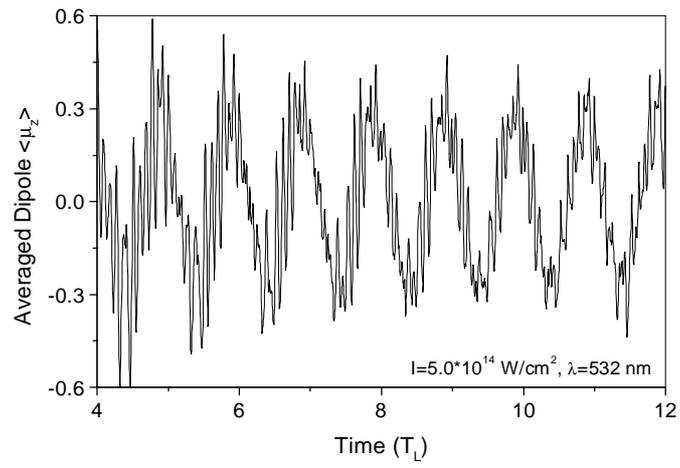
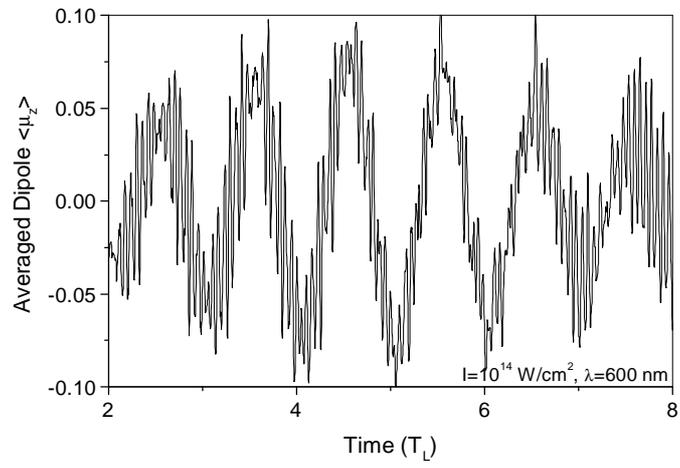
\begin{figure}[tbp]
\caption{Temporal variation of the averaged dipole moment for different
laser parameters. This first one corresponds to the hydrogen atom with
initial energy $E_{0}=-0.5$ $hatree$, the other one corresponds to the
hydrogen molecular ion with initial energy $E_{0}=-1.1034$ $hatree$ and
internuclear distance $R=2.00$ $bohr$.}
\end{figure}
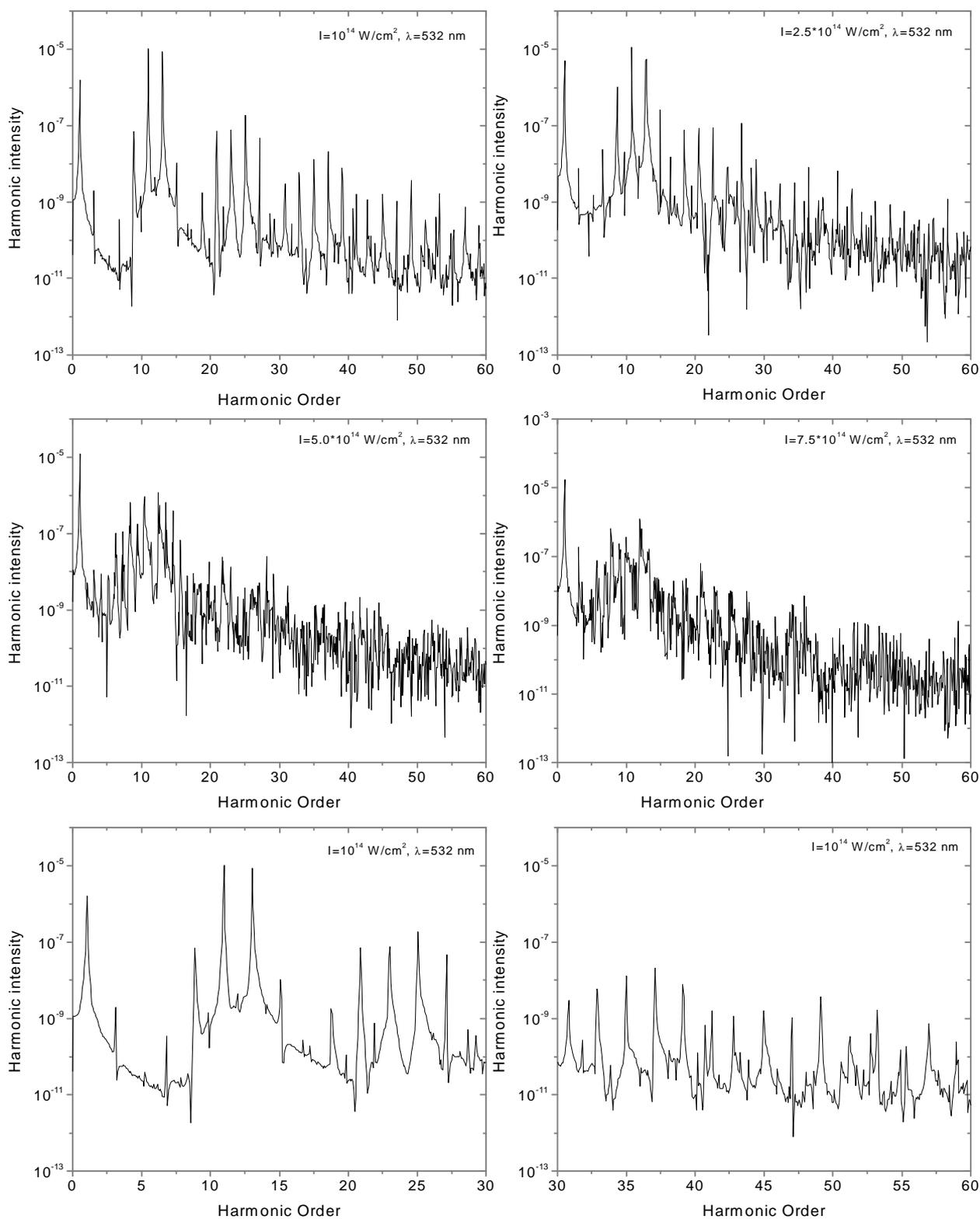
\begin{figure}[tbp]
\caption{Harmonic spectra of the ground-state hydrogen atoms interacting
with different laser pulses, the last two are magnifications of the first
one.}
\end{figure}
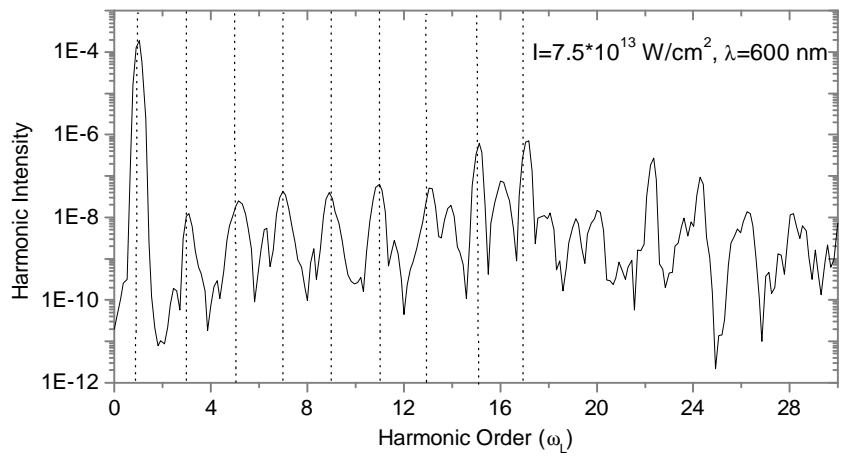
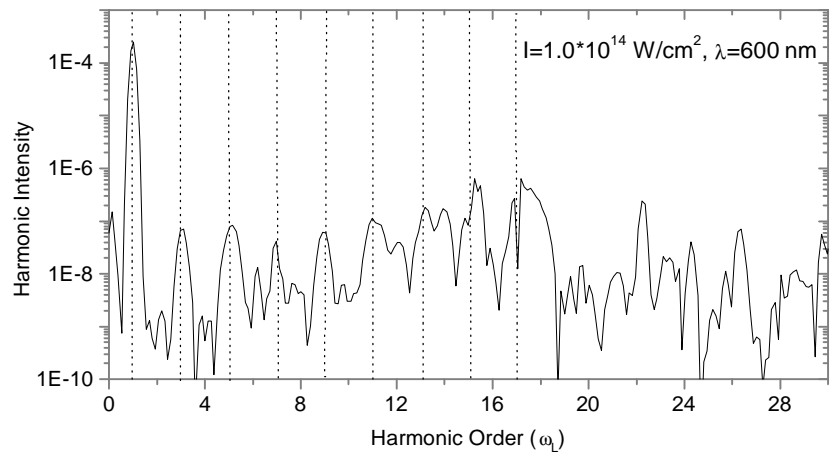
\begin{figure}[tbp]
\caption{Harmonic spectra of the ground-state hydrogen molecular ion
interacting with different laser pulses, with initial energy $E_{0}=-1.1034$ 
$hatree$ and nuclear internuclear distance $R=2.00$ $bohr$.}
\end{figure}

\end{document}